\documentclass[twocolumn,showpacs,preprintnumbers,amsmath,amssymb,aps,prl]{revtex4}
\usepackage{graphicx}% Include figure files
\usepackage{dcolumn}% Align table columns on decimal point
\usepackage{bm}% bold math

\begin{document}
\pagestyle{empty}
\preprint{}

\title{Structural Determinant of Protein Designability}%:\\with Forced Linebreak}% Force line breaks with \\

\author{Jeremy L. England}
\author{Eugene I Shakhnovich}%
 \email{eugene@belok.harvard.edu}
\affiliation{%
Harvard University, Department of Chemistry and Chemical Biology, 12 Oxford Street, Cambridge, Massachusetts 02138
}%

%\affiliation{

\date{\today}% It is always \today, today,

\begin{abstract}
Here we present an approximate analytical theory for the relationship 
between a protein structure's contact matrix and the shape of its energy 
spectrum in amino acid sequence space. We demonstrate a dependence
of the number of sequences of low energy in a structure on 
the eigenvalues of the structure's contact matrix, and then use a Monte Carlo simulation to test the 
applicability of this analytical result to cubic lattice proteins. We find that the lattice
structures with the most low-energy sequences are the same as those predicted
by the theory. We argue that, given
sufficiently strict requirements for foldability, 
these structures are the most designable, and we propose a simple means to test whether
the results in this paper hold true for real proteins.
% \verb+\pacs{#1}+ command.
\end{abstract}

\maketitle

%\section{\label{sec:level1}First-level heading:\protect\\ The line
%break was forced \lowercase{via} \textbackslash\textbackslash}

Successful protein design relies in part on knowledge of how a polypeptide 
chain's native structure varies as a function of its amino acid sequence. Yet even if 
this ``protein folding problem'' were solved in its entirety,
the would-be protein designer would still face the formidable task of 
\textit{finding} those sequences which folded to the target structure he or she
wished to engineer \cite{MAYO_SCI,DES_REV}. 
It is therefore vital to know how many solutions to this search of
sequence space exist
for a given target structure, i.e$\textnormal{.}$ how 
\textit{designable} the target structure is
\cite{FBG,Tang,DES_REV,Kussell:99,Buchler:2000,Koehl:2002}. 
The question of what makes a particular protein 
structure more designable also bears fundamentally on our understanding of how proteins first
evolved. 

Past studies of designability have been limited in large part by their lack of generality.  The previous
contributions in \cite{Tang,Kussell:99}, for example, rely heavily on their study
of two-letter monomer alphabets and Cartesian lattice-polymers, and can therefore
offer no clear implications for the 20-letter, off-lattice world of real proteins \cite{DES_REV}.  In contrast, 
studies like that of Koehl and Levitt \cite{Koehl:2002} come closest to probing real protein designabilities,
but have no theoretical foundations from which to extrapolate beyond their numerical results.  Finally, a number of
investigations have assumed that the distribution of amino acid sequence
energies is either nearly \cite{Tiana:01} 
 or 
totally \cite{Buchler:2000} independent of the structural topology of the target fold, a premise which is 
flatly contradicted by the findings in \cite{Kussell:99}.  The intention of this communication is therefore to identify 
a theoretically motivated, generally applicable quantitative measure of structural topology which we expect to be a good
predictor of designability.

In this Letter, we develop an approximate analytical theory of the spectrum of possible monomer sequence
energies for a heteropolymer in any given conformation.  Our theory leads us to identify a novel topological property
of a conformation, its contact trace, as a determinant of the shape of the conformation's sequence spectrum.  We then 
confirm the predicted usefulness of the contact trace by running Monte Carlo simulations under conditions which are more realistic
than those for which the contact trace was originally identified.  Finally, we use the results of these simulations to 
suggest a connection between the contact trace and designability, and we propose a way to test this hypothesis on
real protein structures.

As a preliminary, we must draw a distinction between what 
we term ``\textit{strong} designability'' and ``\textit{weak} designability''. 
We define a structure's ``strong designability'' in the same 
way that previous studies have defined its ``designability'': 
as the total number of amino acid sequences that 
fold to that structure. 
Thereotical arguments based on the random energy model suggest that 
 in order for a sequence to be foldable, its native state energy must lie
below a certain low-energy threshold $E_c$ \cite{NATUR,B.W.,FGB,Kussell:99,SSK1},
which represents the lowest energy range accessible to misfolded conformations.
We therefore define a structure's 
``weak designability'' to be the 
fraction of its sequences which lie below
a low energy cutoff $E_{c}$.
It is important to note that though 
the strong and weak versions of 
designability are set apart
\textit{a priori}, it is possible 
that they will turn out to be the 
same thing in practice. Indeed, we 
will argue later on that if 
natural thermodynamic and kinetic 
requirements for folding are sufficiently 
stringent, then a structure's strong and 
weak designability 
become essentially indistinguishable.

Our first aim is to determine whether contact 
topology effects a structure's distribution 
of possible sequence energies, and to this end 
we derive a closed-form
sequence partition function for a special 
class of amino acid alphabets.  We begin by considering a 
polymer of $N$ monomers, where each 
monomer can be one of $2M$ different
possible kinds.  
We may construct for any polymer configuration 
an $N\times N$ traceless contact matrix $C$ whereby 
lattice nearest neighbors that are not sequence neigbors are considered
to be in contact \cite{CHPH}.
Next, we represent the amino acid type of 
the $i$th monomer in the chain as 
a $2M$ dimensional unit-vector $\vec{s}^{\textnormal{ }(i)} = [0,\ldots,0,1,0,\ldots,0]$, where the non-zero vector element is the $k$th
element of the vector if the $i$th monomer is of type $k$.

We define the Hamiltonian to be a standard 
nearest-neighbor contact potential, 
i.e$\textnormal{.}$
\begin{equation}
\mathcal{H} = \frac{1}{2}\sum_{i,j}^{N,N}C_{i,j}\vec{s}^{\textnormal{ }(i)}\cdot(B\vec{s}^{\textnormal{ }(j)})
\label{eq:one}
\end{equation}
where $B$ is the $2M\times 2M$ matrix of 
interaction energies for the different 
pairs of monomer types. A closed form expression for the partition function of this Hamiltonian may be obtained if we let $B$ take on the special form
\begin{equation}
B = \begin{bmatrix} V_{1,1} & -V_{1,1} & V_{1,2} & -V_{1,2} &\ldots \\
    -V_{1,1} &  V_{1,1} & -V_{1,2} & V_{1,2}&\ldots\\
    \vdots & \vdots &\vdots &\vdots & \end{bmatrix}
\label{eq:two}
\end{equation}
In other words, $B$ is the direct product of a matrix $V$ with an Ising-like
potential.    Though we have now constrained
the nature of our contact potential, we still retain unlimited
freedom to choose the types of interactions represented in $V$.

Now let us define the $M$-vector $\vec{\sigma}$ through $\sigma_{k+1} = s_{2k+1}-s_{2k+2}$. 
Our Hamiltonian becomes
\begin{equation}
\mathcal{H} = \frac{1}{2}\sum_{i,j}^{N,N}C_{i,j}\vec{\sigma}^{\textnormal{ }(i)}\cdot(V\vec{\sigma}^{\textnormal{ }(j)})
\label{eq:three}
\end{equation}
where $\vec{\sigma}^{(i)}$ is a vector of unit length whose single non-zero element may be either $1$ or $-1$.

At this point we define the $MN\times MN$ block matrix $\mathbf{U}$ through $\mathbf{U}_{M(i-1)+k,M(j-1)+l}=C_{i,j}V_{k,l}$.  In other words,
we turn every element of $C$ into an $M\times M$ block which couples the vectors $\{\vec{\sigma}^{\textnormal{ }(i)}\}$ through
$V$ wherever there is a contact.  Finally, if we write $\pmb\sigma\equiv [\vec{\sigma}^{\textnormal{ }(1)},\ldots,\vec{\sigma}^{\textnormal{ }(N)}]$
then the Hamiltonian can be expressed as
\begin{equation}
\mathcal{H} = \frac{1}{2}\pmb\sigma\cdot(\mathbf{U\pmb\sigma})
\label{eq:four}
\end{equation}
This form of the Hamiltonian will allow us to calculate the sequence space 
partition function, for any contact map and
any potential matrix $V$.  To do so, we employ a continuous-spin approximation, allowing each vector
$\vec{\sigma}^{\textnormal{ }(i)}$ to swing anywhere on the $M$-dimensional unit sphere instead of restricting it
to one of the $2M$ available unit-integer lattice points.  This approximation not only smears the discrete sequence
spectrum into a continuous one, but also distorts the spectral width by altering the relative sizes of $M$ and $M-1$ letter
sequence spaces, thereby preventing the theory from making quantitatively
accurate predictions.  Our assumption is that the model nevertheless retains important information about the \emph{effect} that 
variations in contact topology 
have on the shape of the spectrum.  The partition function now becomes
\begin{equation}
Z(\beta) = \int d^{MN}\pmb\sigma \left(\prod_{i=1}^{N}\delta\left(|\vec{\sigma}^{\textnormal{ }(i)}|-1\right)\right)
\exp\left[-\frac{\beta }{2}\pmb\sigma\cdot(\mathbf{U\pmb\sigma})\right]
\label{eq:five}
\end{equation}
Assuming hereafter that $M\gg 1$, we have
\begin{equation}
d^{M}\vec{\sigma} \exp\left[-\frac{(M-1)}{2}|\vec{\sigma}|^{2}\right] \simeq d|\vec{\sigma}|\textnormal{ }\delta(|\vec{\sigma}|-1)
\label{eq:six}
\end{equation}
Defining $\mathbf{M} = (M-1)\mathbf{I}$, where $\mathbf{I}$ is the $MN\times MN$ identity matrix, and $\mathbf{u} = \mathbf{M}^{-1}\mathbf{U}$ we may write
\begin{equation}
Z(\beta) = \int d^{MN}\pmb\sigma\exp\left[-\frac{1}{2}\pmb\sigma\cdot(\mathbf{M}+\beta\mathbf{U})\pmb\sigma\right]
\label{eq:seven}
\end{equation}
and, since $\mathbf{U}$ is a real symmetric matrix, we finally obtain
\begin{equation}
\frac{Z(\beta)}{Z(0)} \equiv z(\beta)= \sqrt{\frac{\det[\mathbf{M}]}{\det[\mathbf{M+\beta\mathbf{U}}]}}=\frac{1}{\sqrt{\det[\mathbf{I}+
\beta\mathbf{u}]}}
\label{eq:eight}
\end{equation}
If we use $\det\mathbf{A} = \exp\textnormal{Tr}\ln\mathbf{A}$ to expand $\ln z$ to
$\mathcal{O}(\beta^{2})$ and inverse-Laplace transform the resulting approximation of the partition function to obtain a distribution
of sequence energies $n(E)$, we get
\begin{equation}
\begin{aligned}
n(E) \approx \frac{1}{2\pi}\int e^{i\beta E} z(i\beta)d\beta \simeq\exp\left[-\frac{E^{2}}{\textnormal{Tr }\mathbf{u}^{2}}\right] \\
\simeq
\exp\left[-\frac{E^{2}}{2N_{C}\sigma_{B}^{2}}\right]
\end{aligned}\label{eq:REM}
\end{equation}
%\begin{equation}
%\begin{aligned}
%\textnormal{Tr }\mathbf{u}^{2}=(\textnormal{Tr }C^{2})(\textnormal{Tr } v^{2})\\
%=(\sum_{ij}C_{ij}C_{ji})\left(\sum_{ij}\langle j|v|i\rangle \langle i|v|j\rangle\right) \\
%=2N_{C}\left(\sum_{ij}|\langle j|v|i\rangle|^{2}\right)\\
%\simeq 2N_{C}\left(\frac{1}{4M^{2}}\sum_{ij}|\langle j|B|i\rangle|^{2}\right)=2N_{C}\sigma_{B}^{2}
%\end{aligned}\label{eq:ten}
%\end{equation}
where $N_{C}$ is the number of contacts and $\sigma_{B}^{2}$ 
is the variance of the interaction energies in $B$.
A similar result was obtained in \cite{DES_REV} by a straightforward expansion
of the sequence partition function.  
Equation (\ref{eq:REM}) says that
the Gaussian approximation of the density 
of states leads to a na\"ive sequence space 
random energy model (REM) for $n(E)$.  
It is this REM picture, in which all 
structures with the same number of contacts
have identical sequence spectra,  
which has been used implicitly in \cite{Buchler:2000}.  
We demonstrate below, however, that the consideration of 
higher order terms in the expansion of $\ln z$ can
have a profound impact on how we understand designability.

Let us now consider the free energy $F=-\frac{1}{\beta}\ln z$ of 
our sequence space partition sum.  Defining 
the matrix $v=\frac{1}{M-1}V$ and expanding about high ``design 
temperature'' (i.e. requiring that$|\beta\lambda_{i}| < 1$, for 
all $\lambda_{i}$ which are eigenvalues of $\mathbf{u}$), we get
\begin{equation}
\begin{aligned}
F = -\frac{\beta}{4}(\textnormal{Tr }v^{2})(\textnormal{Tr }C^{2})
+\frac{\beta^{2}}{6}(\textnormal{Tr }v^{3})(\textnormal{Tr }C^{3})\\ -\frac{\beta^{3}}{8}(\textnormal{Tr }v^{4})(\textnormal{Tr }C^{4}) +
\mathcal{O}(\beta^{4})
\end{aligned}\label{eq:series}
\end{equation}
Those structures which minimize $F$ will 
be the ones with the greatest number of 
amino acid sequences
that have low energy when threaded onto 
that structure.  Minimization of $F$ is 
therefore a means to maximize
weak designability, which we recall is 
determined by the fraction of low-energy 
sequences in a structure's sequence spectrum. 
Thus, we now consider which choices of the 
contact matrix $C$ would serve to make the 
free energy $F$ as negative 
as possible.  Because $\textnormal{Tr }C^{n}$ 
is equal to the number of $n$-step paths along 
the contact map which return to their
starting place, we know that all such contact 
traces must be positive.  Thus, the exact 
behavior of the series in (\ref{eq:series}) 
will hinge on whether
the largest eigenvalues of $v$ are positive 
or negative. 

\begin{figure}[t]
%\begin{figure}[p]
%\resizebox{\columnwidth}{!}{\includegraphics{figures/thermo_prl2.gif}}
\resizebox{190pt}{!}{\includegraphics{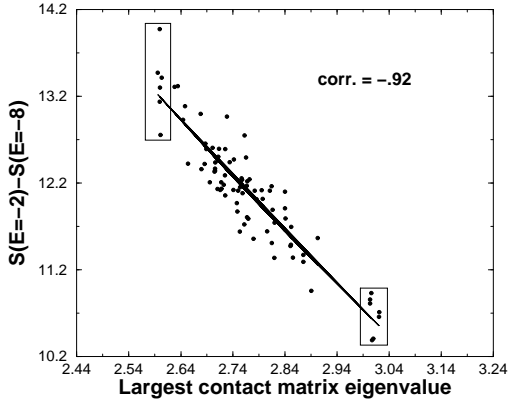}}
\caption{\label{fig:corr} The difference in sequence space entropy between
an energy near the peak of all structural sequence spectra ($E=-2$) and one in the lower tail
of all spectra ($E=-8$) as a function of the contact trace, measured here by the largest eigenvalue of
the structure's contact matrix (which follows from $\textnormal{Tr }C^{n}=\sum\lambda_{i}^{n}$).  Each point was generated from data collected while slowly annealing a Monte Carlo sequence design
simulation from high temperature ($T=2$) to low ($T=.2$), with $10^{7}$ Monte Carlo steps taken at each temperature.  The boxed points
correspond to structures which
were chosen by hand so as to ensure that the extrema of the range of possible eigenvalues were represented.  All other structures 
were chosen randomly.}
\end{figure}

\begin{figure}[t]
%\begin{figure}[p]
%\resizebox{\columnwidth}{!}{\includegraphics{figures/thermo_prl2.eps}}
\resizebox{190pt}{!}{\includegraphics{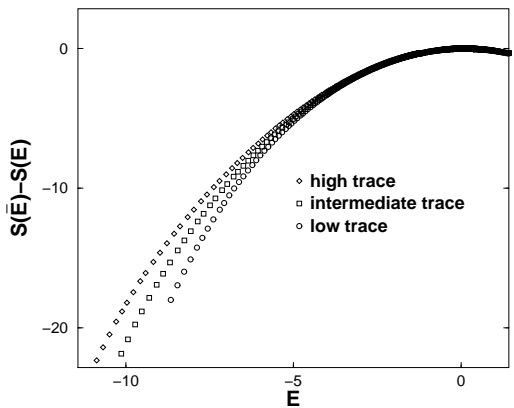}}
\caption{\label{fig:ne} The change in sequence space entropy from energy $\langle E\rangle (T=\infty)=\overline{E}$ to energy $E$
for three structures with largest contact matrix eigenvalues of $3.02$ (high trace), $2.78$ (intermediate trace), and $2.60$ (low trace). }
\end{figure}

For either type of potential matrix $v$, however, we expect that there will
be some positive correlation between the trace of an even power of a structure's
contact matrix and the number of low energy monomer sequences in that structure.  Furthermore, the dependence of the free energy expansion in
(\ref{eq:series}) on such coarse quantities as the traces of powers of $v$ suggests that the impact of the contact matrix on the spectrum of sequence energies should be relatively insensitive to the detailed features of the potential.  We therefore determined to empirically test 
whether the above results remained valid for a discrete monomer alphabet which violated
the special form of the potential assumed in (\ref{eq:two}).  
We first calculated
the contact matrices for all $103346$ different 
compact conformations of 27-mers on a cubic lattice 
\cite{CHPH}.  
Next, we calculated 
$\langle E\rangle$ vs. $T$ 
annealing curves for random starting sequences 
on different structures for a standard 
Monte Carlo search of sequence space 
with a move set containing composition-preserving 
two-monomer and three-monomer permutations.  The energy of each sequence was determined using
a potential set given by Table 6 of \cite{MJ}. This
set of interactions, where average interactions are subtracted out, is one of the most diverse potentials possible for a 20-letter alphabet, and therefore provides the most general empirical test of the predicted relationship between sequence energies and contact topology.
From the annealing curves, we then calculated the entropy in sequence space $S(E)$ according 
to the prescription given by  eq$\textnormal{.}$ (11) of \cite{DES_REV}.

Figure \ref{fig:corr} plots the sequence space entropy difference between low and 
near-modal energy versus
the largest eigenvalue of the structure's 
contact matrix for $86$ randomly-selected lattice
structures.  
As predicted, the entropy difference between
the peak and the left tail decreased as the 
largest contact matrix eigenvalue increased 
(corr. $= -0.92$), indicating that more sequences 
have low energy in high trace structures. 
Figure \ref{fig:ne} illustrates that the 
effect observed in Fig$\textnormal{.}$ \ref{fig:corr} results from 
global differences in the shapes of 
the sequence spectra of high trace and 
low trace structures.  The higher the 
contact trace, the 
more gradually
the number of sequences falls off as 
energy decreases, and therefore the greater 
the relative number of sequences of low energy.
Clearly, the contact trace of the target 
structure controls how low in energy a Monte Carlo 
sequence optimization algorithm running 
at fixed temperature $T_{des}$ will be able to search.  
The greater the contact trace, the larger the $S(E)$ at low energies, i.e. 
the greater the weak designability.

%\begin{figure}[t]
%%\begin{figure}[p]
%%\resizebox{\columnwidth}{!}{\includegraphics{figures/thermo_prl2.eps}}
%\resizebox{\columnwidth}{!}{\includegraphics{Fig3.eps}}
%\caption{\label{fig:str} Representative $27$-mer lattice structures of maximal and minimal contact trace.}
%\end{figure}
 Interestingly, the most designable 27-mer structures
identified using our maximum eigenvalue determinant are similar to the one
identified in \cite{Tang02} using random
sampling of sequences and a different, ''solvation-like''  Miyazawa-
Jernigan potential. This attests to a generality of our proposed structural
determinant of designability with respect to potentials.

%Interestingly, the contact trace of a structure appears 
%to be related to its regularity and symmetry.
%For a fixed 
%number of contacts, a structure maximizes 
%the number of $n$-step closed paths in 
%its contact system by having a contact
%map with a high number of contact loops 
%made from  netted contact strands.  In 
%contrast, minimization of the contact trace requires
%the contact system to be as un-looped 
%as possible, with long contact strands 
%branching and dead-ending at surface sites. 
% As Figure
%\ref{fig:str} shows, these different constraints 
%lead to remarkable differences in the regularity and
% symmetry of high and low trace
%structures. This sheds light on the earlier observation \cite{Tang}
%that highly designable structures tend to be more symmetrical.

\begin{figure}[t]
%\begin{figure}[p]
%\resizebox{\columnwidth}{!}{\includegraphics{figures/thermo_prl2.eps}}
\resizebox{190pt}{!}{\includegraphics{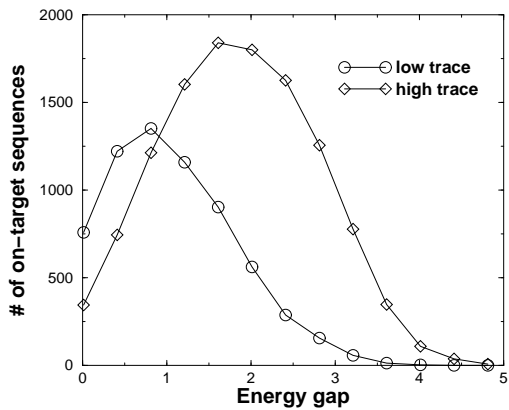}}
\caption{\label{fig:gaps} We designed sequences on two different target 
structures in Monte Carlo searches  of $2\times 10^{7}$ steps 
($.2<T\leq .4$), sampled at every one sequence in $10^{4}$.  Here, 
we plot the number of sequences whose unique compact 
ground states were the structure on which they
were designed, as a function of the difference 
in energy between the sequence's ground state 
and the lowest-lying excited state
which shares fewer than $25$\% of its 
contacts with the ground state.}
\end{figure}

Structures of high contact trace are weakly
 designable, but are they strongly designable?  
 In order to address this question,
we examined the stability of sequences designed 
on two structures of maximal and minimal contact trace. 
For each target structure, we
determined how many of its designed 
sequences were ``on-target'', that is, 
had the target structure as their unique energetic ground 
state determined over all compact conformations, 
and calculated the gap in energy between that of 
the ground state and those of the lowest-energy structures with low structural similarity
to the ground state conformation.  
We found that while only $12$\% of 
the sequences designed for the low trace 
structure were on-target, $24$\%
were in the case of the high trace 
structure.  Furthermore, as Figure \ref{fig:gaps} 
illustrates, the sequences
designed successfully on the high trace structure tended on 
average to have larger energy gaps than their 
low trace cousins, consistent with earlier observations that low
energy in the native conformation also leads on-average to a larger
gap and greater
stability \cite{PNAS,TRAPS} 

Fig$\textnormal{.}$ \ref{fig:gaps} therefore gives us 
a means to unify strong and weak designability.  
Past studies have suggested that in 
order for a protein sequence to fold 
rapidly to its native state, it must have a 
larger-than-average energy gap \cite{GSW,SSK1,Raleigh:2001}.
Fig$\textnormal{.}$ \ref{fig:gaps} demonstrates that if
the conditions for folding stably and 
quickly in nature demand sufficiently 
high energy gaps, weak designability
will be one and the same with 
strong designability, since only 
by having very low energy in its 
target structure will a sequence have an 
appreciable chance of exhibiting 
the gaps which are thermodynamically 
and kinetically required by the natural environment.  
We speculate that protein evolution 
under such conditions would lead 
to the emergence of natural protein
folds with near-optimal contact traces.  

We have presented an analytical 
theory which identifies the first instance of a generally applicable, 
well-defined, numerical measure of a protein structure's topology which 
is expected to correlate with the structure's designability.
Using a Monte-Carlo search in the sequence space of a lattice model
with $20$-letter energetics, we have shown that the
implications of the theory extend beyond the special assumptions under which 
they were first developed.  The finding that higher contact trace
may lead to greater potential for thermal stability
leads us to hypothesize that thermophilic organisms, whose natural environment
makes uncommonly strict demands for protein stability, exhibit
a genomic bias towards folds of higher contact trace.  We have recently
found that this bias exists (unpublished results), providing further
encouragement that the contact trace
may yet offer new insight into the laws governing structural diversity in natural proteins.

We thank Edo Kussel, William Chen, Nikolay Dokholyan, and Lewyn Li 
for valuable discussions, 
and NIH and Pfizer Inc. for support.

\bibliography{sglit}% Produces the bibliography via BibTeX.

\end{document}